\documentclass[twocolumn,twocolappendix]{aastex63}
\usepackage{graphicx}

\shorttitle{Excitation of kink oscillations}
\shortauthors{Yu}

\begin{document}
\title{Effects of external flow on resonant absorption of coronal loop kink oscillations driven by an external fast wave: Selective excitation problem}

\author[0000-0003-1459-3057]{D. J. \surname{Yu}}
\email{djyu79@gmail.com}
\affiliation{Department of Astronomy and Space Science, Kyung Hee University,
 1732, Deogyeong-daero, Yongin, Gyeonggi 17104, Republic of Korea}

\begin{abstract}
Resonant absorption is considered as a crucial mechanism for the damping of the coronal loop oscillations and plasma heating. We study resonant absorption of the coronal loop kink oscillations excited by such external drivers as flares on the assumption that there is an intermediate shear flow region surrounding the loop. We find that for long coronal loops resonant absorption can be highly enhanced or reduced depending sensitively on the magnitude and direction of the flow and the spatial extent of the flow region when the transitional layer is thin. For short coronal loops, high flow speed and thick transitional layer are needed to have a substantial resonant absorption. We provide a potential picture to explain the results where the external Alfv\'{e}n speed and phase speed of the wave are important parameters. These results imply that the transport of the external wave energy into the loop is significantly changed by the shear flow region, which may cause the selective excitation of the coronal loop oscillations.
\end{abstract}


\section{Introduction}
\label{sec1}
Resonant absorption of magnetohydrodynamic (MHD) waves is considered as a crucial mechanism for plasma heating in solar corona and a useful tool for the coronal seismology~\citep[e.g.][]{VanDoorsselaere2020,Nakariakov2020}. The coronal plasmas usually have loop-like structures where MHD wave modes like kink (transverse) and sausage oscillations are frequently observed~\citep{Li2020,Nakariakov2020,Wang2021}.
The loop oscillations are in general excited by internal or external driving sources~\citep{Nakariakov2020}.
{The former includes foot-point motions of the loop~\citep{Berghmans1997,Tirry1997,DeGroof2000,DeGroof2002} and colliding flows in the loop~\citep{Antolin2018}. For the latter, the vortex shedding~\citep{Nakariakov2009,Karampelas2020,Karampelas2021} and magnetic energy release~\citep{Russell2015,Sarkar2017} have been discussed.}
The transient event like flares can perturb its environment, which plays a role of wave source that may propagate to nearby loops and excite the kink oscillations~{\citep[e.g.][]{Uralov2003,Terradas2009,Aschwanden2011,Selwa2011,Ofman2015}}. But, in general, the Alfv\'{e}n speed outside the loops, which corresponds to the cutoff phase speed, is higher than the inside, not allowing in principle the energy transport to the loops. It is only possible if the distance between them is sufficiently short or the phase speed of the wave is higher than the cutoff~\citep[e.g.][]{Verwichte2006,Terradas2009}.

These cognitions are based on no background flows. The observations of omnipresent shear flow~\citep[e.g.][]{Morgan2018,Tian2021} raise the question about the wave propagation properties in the presence of the background flow.
{Spectroscopic investigations of Dopple shift show red/blue shift of the flow speed from a few $kms^{-1}$ to $\sim100kms^{-1}$ depending on the emission lines (temperature) and the regions considered~\citep[e.g.][]{Dammasch1999,Peter1999,Xia2003,Tu2005,Dadashi2011,Mcintosh2012,Wang2013,Young2015}. The flow speed may reach $~200kms^{-1}$ in active region boundary, which is mostly in the range of $50-150kms^{-1}$ for upflows~\citep[e.g.][]{Kamio2011,Tian2011,Ugarte-Urra2011,Boutry2012,Mcintosh2012,Harra2012}. It seems that flows with speeds of several tens of $kms^{-1}$ occur frequently in the solar atmosphere.}

It is more difficult to understand the wave properties in the shear flow region~\citep{Terradas2011,Ofman2012,Li2014,DeMoortel2015,Pant2015,Provornikova2018,Bahari2020,Cho2020,Yu2020}. It was previously shown that resonant absorption may change significantly in the presence of background flow inside the loops~\citep{Goossens1992,Ruderman2010,Terradas2010,Soler2011,Ruderman2019,Soler2019,Sadeghi2021}, while the effects of the flow outside the loops on resonant absorption is not well-known. The excitation mechanism of the loop oscillations is also not fully understood yet and one of the puzzling features is their selective excitations: why do certain loops show oscillations whereas their neighboring ones do not? The external driving is thought to not explain the selectivity of the excitation~\citep{Nakariakov2020}. This consensus is also based on the assumption of no external background flow. Therefore, it is necessary to understand the effects of background flow outside the loop on the excitation and relevant resonant absorption of the loop oscillations.

In this work, we investigate the influence of the external background flows on resonant absorption of the {fundamental standing} kink modes. We consider that an external wave source due to transient events like flares is excited at some external site and propagates, through the inhomogeneous flow region, to a coronal loop, exciting the kink oscillations which then undergo damping by resonant absorption. {As we are concerned with the fundamental standing kink oscillations, we consider as an incident wave the kink modes with the frequency of the fundamental mode. In general, the wave source consists of multiple harmonic components of azimuthal wave numbers, axial wavelengths, and frequencies~\citep[see, e.g.,][and references therein]{Terradas2009}.}

\section{Model and Method}
\label{sec2}
We model a coronal loop as an infinitely long, straight, and axisymmetric cylindrical flux tube. We restrict inhomogeneity to radial direction and assume that the background magnetic field and flow are in axial direction:~$\textbf{\textsl{B}}=(0,0,B_{0})$ and $\textbf{\textsl{U}}=(0,0,U_{0}(r))$. Linearization of ideal MHD equations and Fourier transform with the factor $\exp[i(-\omega t+k_zz+m\phi)]$ {lead to}
the wave equation for the total pressure perturbation $P$
~\citep[e.g.][]{Goossens1992,Soler2019}
\begingroup
\begin{eqnarray}
\frac{d^2 P}{d r^2}+\bigg[\frac{1}{r}-&&\frac{1}{D}\frac{dD}{dr}
\bigg]\frac{d P}{d r}+\bigg[\frac{\Omega^2-\omega_A^2}{v_A^2}-\frac{m^2}{r^2}\bigg]P=0~~\label{eq:1}
\end{eqnarray}\endgroup
for coronal plasmas (low-beta approximation), where $D(r)=\rho_0(r)(\Omega^2(r)-\omega_A^2(r))$, $\Omega(r)=\omega-k_zU_0(r)$, $\omega_A^2(r)=k_z^2v_A^2(r)$, $v_A^2(r)=B_{0}^2/\mu_0\rho_0(r)$, and $\mu_0$ is the magnetic permeability. The MHD wave mode is characterized by the wave frequency $\omega$, the axial wave number $k_z$, and the azimuthal wave number $m$.

In our model for cold plasmas, two parameters are important: density and background flow. To describe the external wave source, in addition to the typical density profile for the coronal loops, we assume a source region with the density $\rho_s$:
\begingroup
\begin{eqnarray}
 \rho_0(r)= \left\{ \begin{array}{ll}
         \rho_i & \mbox{ if $r\leq R_0-l$ }\\
         \rho_a(r) & \mbox{ if $R_0-l<r<R_0+l$ }\\
         \rho_e & \mbox{ if  $R_0+l\leq r\leq R$}\\
         \rho_s & \mbox{ if  $r> R$}
        \end{array} \right.,\label{eq:2}
\end{eqnarray}\endgroup where $R_0$ is the loop radius and $R=R_0+l+R_u$. Here $R_u$ represents the {spatial  extent} of the shear flow region (see Eq.~(\ref{eq:3})) and $l$ is the half thickness of the transitional layer. We consider a linear density profile for the transitional layer:
$\rho_a(r)=(\rho_i-\rho_e)(R_0-r)/2l+(\rho_i+\rho_e)/2$. For the wave generated at the source region to be a propagating mode, the density  $\rho_s$ needs to satisfy the condition $(\omega/k_z)^2>v_{As}^2=B_0^2/(\mu_0\rho_s)$. We set $\rho_s=\rho_i$ to minimize the interference effects induced by the density difference between the wave source region ($\rho_s$) and the inside of the loop ($\rho_i$).
 The background flow has the following structure:
\begingroup
\begin{eqnarray}
 U_0(r)&=&\left\{ \begin{array}{ll}
         0 & \mbox{ if $r< R_0+l$ }\\
         U & \mbox{ if $R_0+l\leq r\leq R$ }\\
         0 & \mbox{ if  $r> R$}
        \end{array} \right.,\label{eq:3}
\end{eqnarray}\endgroup
where $U$ is a constant.
We consider that an external fast MHD wave is incident on a coronal loop with the frequency fixed to the fundamental mode of the standing kink wave, $\omega_k=k_zB_0\sqrt{2/(\rho_i+\rho_e)\mu_0}$ with $k_z=\pi/L$, where ${v_{Ai(e)}=B_0/\sqrt{\mu_0\rho_{i(e)}}}$ and $L$ is the loop length.

The wave source region ($r>R$) is described as a homogeneous medium. The incident and scattered wave functions for $r>R$ can be describe by Bessel ($J_m$) and Hankel ($H_m^{(1)}$) functions of first kind with the scattering coefficient $r_m$~\citep{Yu2016b,Yu2019}:
\begingroup
\begin{eqnarray}
P(r,\phi)&=&\sum_m a_m e^{i m\phi} \big[J_m[k(r-R)+c]\nonumber\\
&&+r_m(R)H_m^{(1)}[k(r-R)+c]\big],~~~~r> R,\label{eq:4}
\end{eqnarray}\endgroup
where $k(=\sqrt{(\omega^2/v_{As}^2)-k_z^2})$ is the radial wave number for $r>R$, $c$ is a constant identical to $kR$, and $a_m$ is an $m$-dependent constant describing the shape of the incident wave. {For example,} when a plane wave is incident to the $x$ direction, $a_m=i^m$~\citep{Stratton2007}. {Here we consider two kink modes ($m=\pm1$) which have equal contribution to resonant absorption~\citep[][]{Yu2016b,Yu2019}.}

We use the invariant imbedding method (IIM)~\citep{Klyatskin2005,Yu2016b,Yu2019} to solve Eq.~(\ref{eq:1}).
Applying IIM to Eq.~(\ref{eq:1}) with matching (boundary) conditions at $r=R$ (Eq.~(\ref{eq:4})), we obtain a differential equation for $r_m$ (see, for detailed derivations,~\cite{Yu2016b}):
\begingroup\small
\begin{eqnarray}
&&\frac{dr_m(r)}{dr}=\frac{k}{H_m}\frac{D(r)}{D_1}[J_m'+r_m(r){H_m^{(1)}}']\nonumber\\
&&+k\bigg[\frac{D(r)}{D_1}\frac{{H_m^{(1)}}'}{H_m^{(1)}}+\frac{1}{k r}\bigg]
\frac{[J'_m+r_m(r){H_m^{(1)}}'][J_m+r_m(r)H_m^{(1)}]}{J'_mH_m^{(1)}-{H_m^{(1)}}'J_m}\nonumber\\
&&+k\bigg[1
-\frac{D_1}{D(r)}\frac{m^2}{k^2 r^2}
\bigg]\frac{[J_m+r_m(r)H_m^{(1)}]^2}{J'_m H_m^{(1)}-{H_m^{(1)}}' J_m},\label{eq:5}
\end{eqnarray}\endgroup
where $D_1=D(r>R)$, $J_m=J_m(c)$, $H_m^{(1)}=H_m^{(1)}(c)$, and prime means $dy(x)/dx$ for $y(x)$. With the initial conditions $r_m(0)=0$, Eq.~(\ref{eq:5}) is integrated from $0$ to $R$, providing $r_m$ at $r=R$~\citep{Yu2016b,Yu2019}.

There appear two singularities in Eq.~(\ref{eq:5}). One is due to wave resonance in Alfv\'{e}n continuum (resonant absorption): $D(r)=0$. To avoid this singularity, we set $\omega=\omega_k+i\omega_i$ with a small $\omega_i(=10^{-8}s^{-1})$, not affecting the results. The other singularity arises from the cylindrical geometry at $r=0$ (see Eq.~(\ref{eq:5})), which is avoided by choosing the plasma parameters at $r/R_0=\delta$ as those for $r/R_0<\delta$. We use $\delta=10^{-6}$ to not affect the results~\citep{Yu2016b,Yu2019}.

Focusing on the kink modes ($m=\pm1$), we may define the absorption coefficient $A$ as~\citep{Yu2016b,Yu2019}
\begingroup
\begin{eqnarray}
A=-\sum_{m=\pm1}\textmd{Re}(r_m+|r_m|^2).\label{eq:6}
\end{eqnarray}\endgroup
In this work, we focus on the absorption by resonant absorption.
\begin{figure}[]
\includegraphics[width=.4\textwidth]{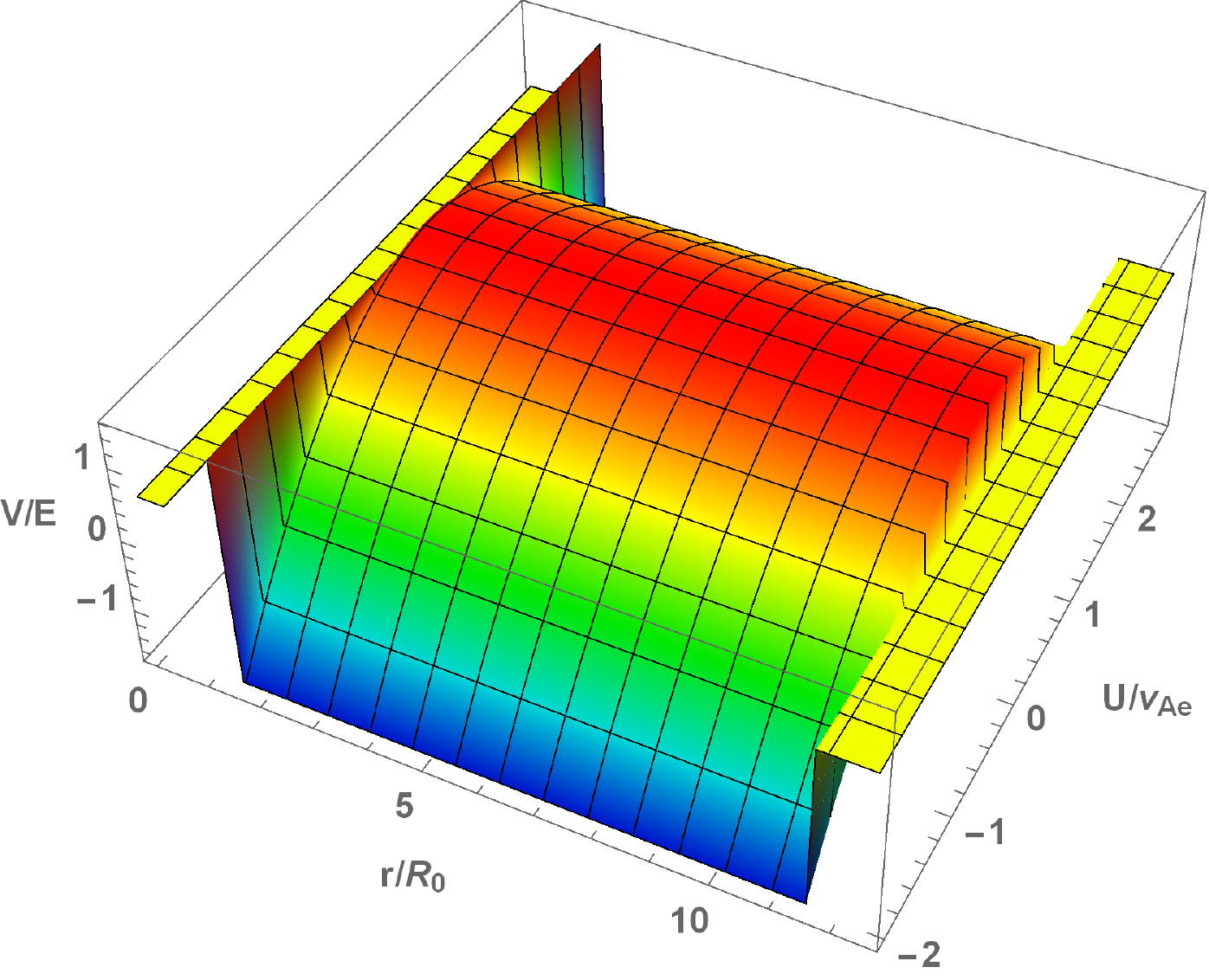}
\caption{\label{f1} A profile of the potential $\delta{V}=V/E$ as functions of $r/R_0$ and $U/v_{Ae}$ when $\omega=\omega_k$, $\delta\rho(=\rho_i/\rho_e)=10$, $l/R_0=0.5$, and $R_u/R_0=10$.}
\end{figure}
\begin{figure}[]
\includegraphics[width=0.5\textwidth]{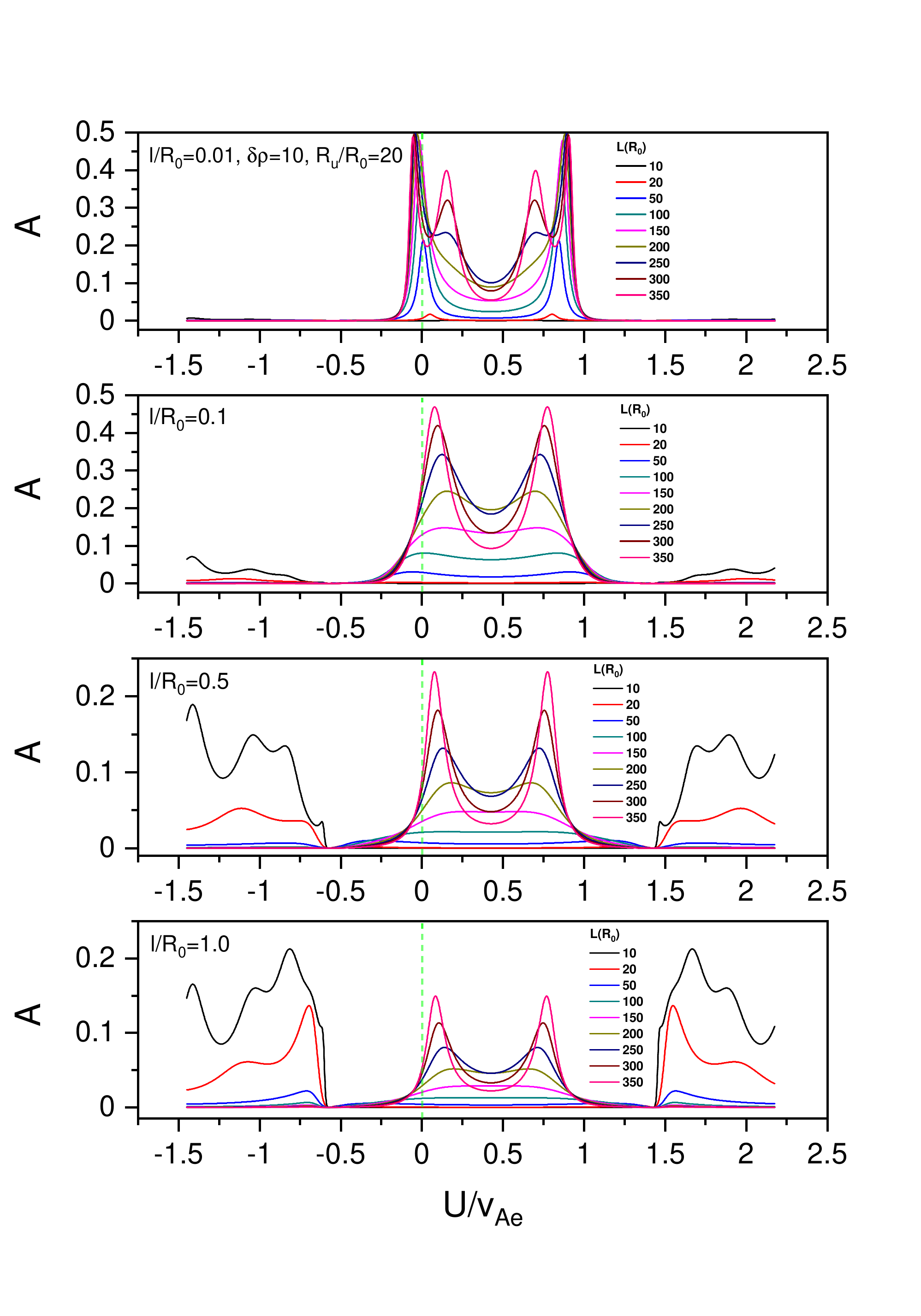}
\caption{\label{f2}Absorption coefficient $A$ vs.~the flow speed ${U/v_{Ae}}$ when $\omega=\omega_k$, $\delta\rho$=10, $R_u/R_0$=20. From top to bottom, ${l/R_0}$=0.01, 0.1, 0.5, and 1.0, and at each panel ${L/R_0}$ changes from 10 to 350. The curves have symmetry with respect to $v_k/v_{Ae}=0.4264$ and $U_{\pm c}/v_{Ae}={v}_k/v_{Ae}\pm1=-0.5736, 1.4264$. The vertical dashed (light green) lines indicate $U=0$.}
\end{figure}
\section{Results}
\label{sec3}
{It is necessary to figure out} the characteristics of the wave equation before showing the results.
 In Eq.~(\ref{eq:1}), the term $(dD/dr)/D$ in the second part  plays a minor role except near the resonance point. As we are concerned with the effect of the flow region ($R_0+l<r<R$), we may use the analogy of the potential view of Schr\"{o}dinger equation. We may define the normalized potential as~\cite[e.g.][]{Yu2010}
\begin{eqnarray}
\delta{V}=\frac{V}{E}=1-\bigg[\frac{(\omega-U_0(r)k_z)^2}{v_A^2(r)}-k_z^2\bigg]\frac{v_{As}^2}{\omega^2}.\label{eq:7}
\end{eqnarray}
\begin{figure}[h]
\includegraphics[width=0.5\textwidth]{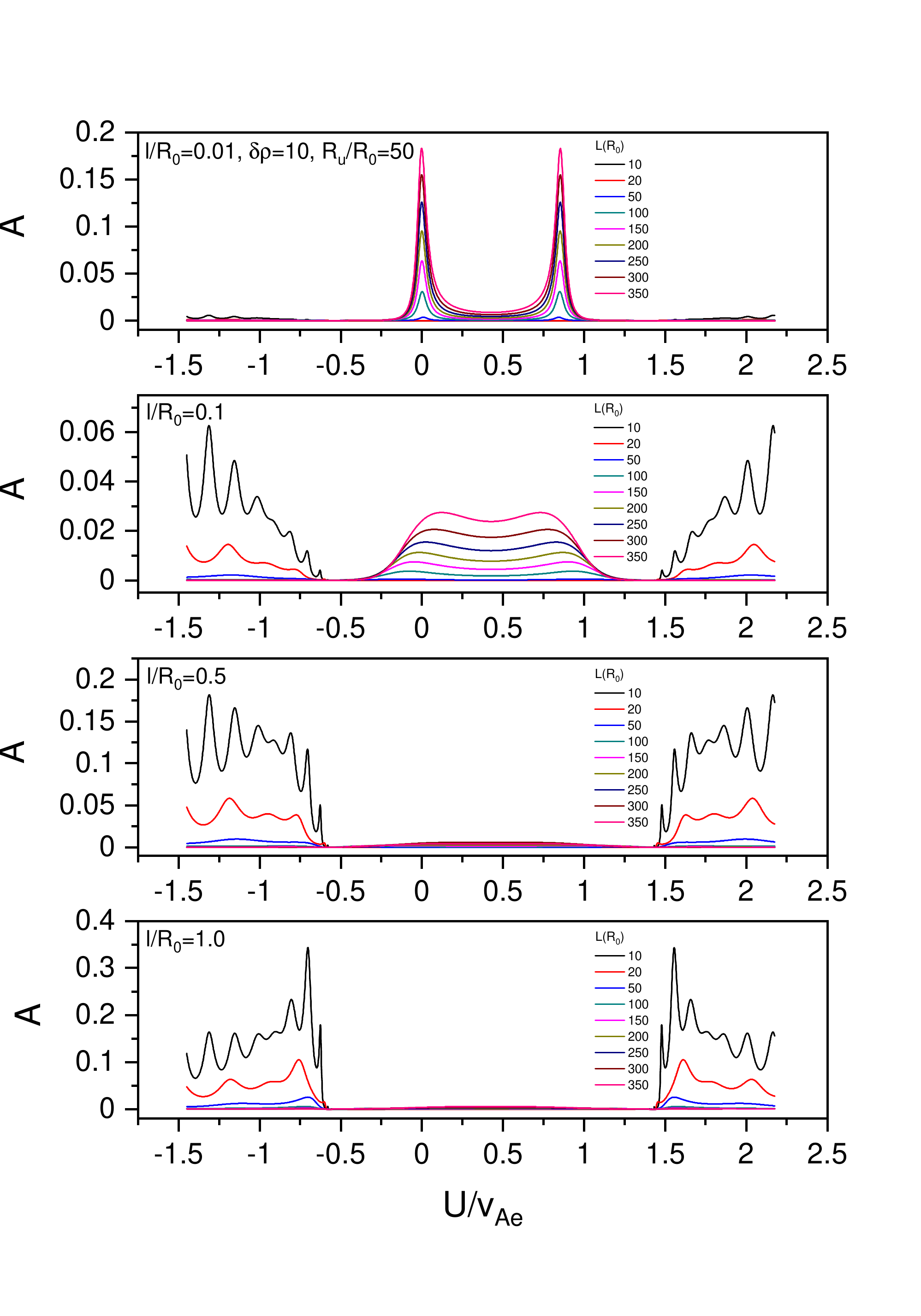}
\caption{\label{f3} $A$ vs. ${U/v_{Ae}}$ when $R_u/R_0$=50 and other parameters are the same in Fig.~\ref{f2}.}
\end{figure}
\begin{figure}[]
\includegraphics[width=0.5\textwidth]{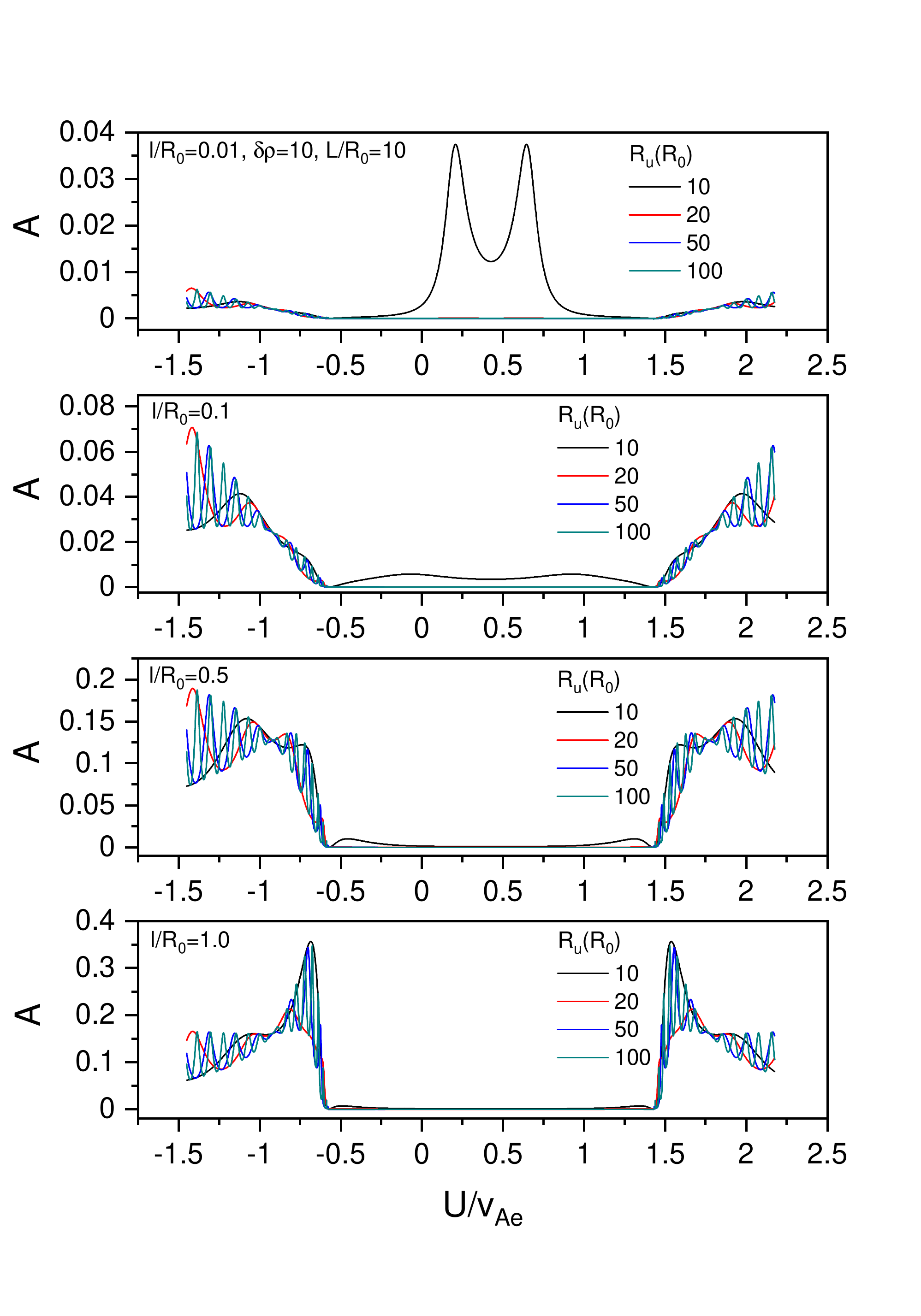}
\caption{\label{f6} $A$ vs. ${U/v_{Ae}}$ when $\omega=\omega_k$, $\delta\rho$=10, and ${L/R_0}$=10, and ${R}_u/R_0$=10, 20, 50, and 100. From top to bottom  ${l/R_0}$=0.01, 0.1, 0.5, and 1.0. }
\end{figure}
Fig.~\ref{f1} shows the potential $\delta{V}$ as functions of $r/R_0$ and $U/v_{Ae}$ when $\delta\rho(=\rho_i/\rho_e)$=10, $l/R_0$=0.5, $\omega=\omega_k$, $L/R_0=50$, and $R_u/R_0$=10. When $U/v_{Ae}$ is small (large), $\delta V$ manifests a potential barrier (well) in the flow region. A typical feature of the potential is its symmetric behavior with respect to the flow speed $U=v_k(=\omega/k_z)$. Another important feature is that $\delta{V}=1$ is a critical value. {Considering} the wave transmission through the rectangular potential barrier~\citep{Cohen1986}, across the flow speed $U_{\pm c}=v_k\pm v_{Ae}$ ($\delta{V}=1$), it is expected that high transmission with oscillatory behavior may occur and the number of transmission peaks increases as $R_u$ increases. We call the {flow regime} for $\delta{V}>1$ ($U_{-c}<U<U_{+c}$) {regime} I and the other {regime} for $\delta{V}<1$ ($U<U_{-c}$ and $U>U_{+c}$) {regime} II. This potential view is not complete since it can not describe the {transitional layer} $R_0-l<r<R_0+l$, but the above features manifest in the results.

In this work we use the parameters in~\cite{Yu2016b} (case II therein) such that $\rho_e=1.67353\times10^{-12}kgm^{-3}$, $B_0=10^{-3}T$, and $R_0=2\times10^6m$, which gives $v_{Ae}=689.757kms^{-1}$.
In Fig.~\ref{f2} we plot the absorption coefficient $A$ vs.~the flow speed ${U/v_{Ae}}$ where $\omega=\omega_k$, $\delta\rho=10$, ${R}_u/R_0=20$, and ${L/R_0}$ is from 10 to 350. From the top to bottom panels, ${l/R_0}=0.01, 0.1, 0.5$, and 1.0. The absorption has symmetry behavior with respect to $U/v_{Ae}={v}_k/v_{Ae}(=0.4264)$ as expected. In {regime} I ($-0.5736<{U/v_{Ae}}<1.4264$), for small ${l/R_0}$, $A$ becomes enhanced as ${L/R_0}$ increases and is strong when ${L/R_0}$ is large, reaching 0.5 as a maximum value. As ${l/R_0}$ increases $A$ gradually decreases for a given ${L/R_0}$. For ${l/R_0}=0.01$ additional peaks appear for ${L/R_0}>250$, which is due to the combination of resonance and tunneling effects. As ${L/R_0}$ increases all absorption peaks tend to shift away from ${v}_k/v_{Ae}$ and their width becomes narrower, implying that the absorption may change greatly for a small change of the flow speed for long coronal loops.
On the other hand, in {regime} II (${U/v_{Ae}}<-0.5736$, ${U/v_{Ae}}>1.4264$), $A$ behaves the opposite, becoming smaller as ${L/R_0}$ increases. For ${L/R_0}\leq50$ $A$ becomes stronger as ${l/R_0}$ increases and multiple peaks appear as ${L/R_0}$ decreases while $A$ in {regime} I is almost zero.

We show in Fig.~\ref{f3} $A$ vs. ${U/v_{Ae}}$ when ${R}_u/R_0=50$ and other parameters are the same as in Fig.~\ref{f2}. The global features are similar to those in Fig.~\ref{f2}, but $A$ changes more rapidly with the increment of ${l/R_0}$. In {regime} I, $A$ is rapidly reduced due to the weak tunneling effect ($\delta{V}>1$) while $A$ in {regime} II is greatly increased and the number of the peaks are increased as expected from the potential view ($\delta{V}<1$). The additional peaks near ${U}={v}_k$ for ${l/R_0}=0.01$ shown in Fig.~\ref{f2} disappear. {The maximum absorption in {regime} II surpasses the value for ${R}_u/R_0$=20 when ${l/R_0}$ is over 0.5. The position of the highest absorption peak is shown to shift toward ${U}_{\pm c}$ as ${l/R_0}$ increases. We do the same calculations for the absorption coefficient by changing $\delta\rho$. It is found that the change of $\delta\rho$ hardly affects the global behavior of the absorption.}

\begin{figure}[]
\includegraphics[width=0.5\textwidth]{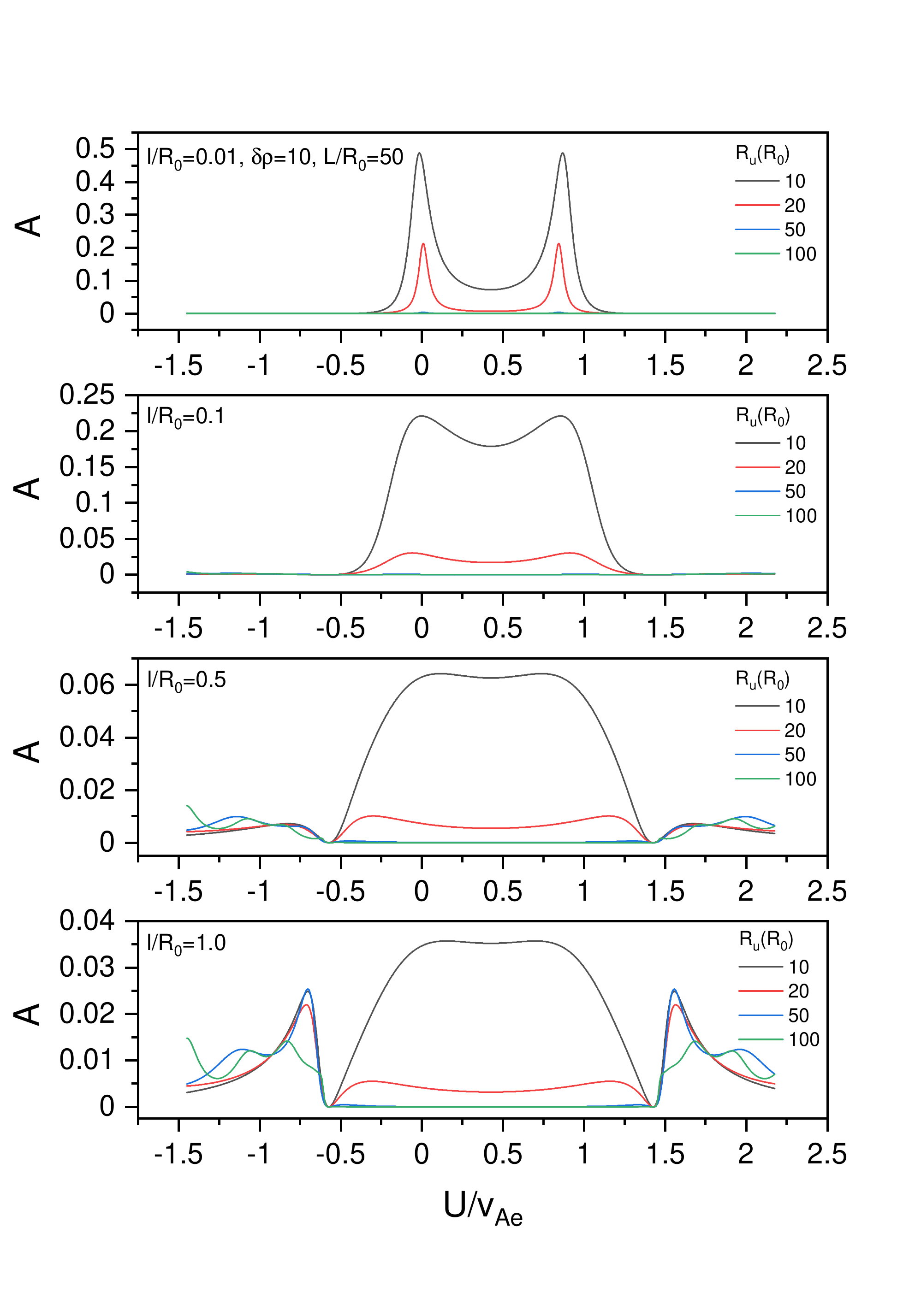}
\caption{\label{f7} Similar to Fig.~\ref{f6} except ${L/R_0}=50$. From top to bottom $l/R_0$=0.01, 0.1, 0.5, and 1.0. }
\end{figure}
\begin{figure}[]
\includegraphics[width=0.5\textwidth]{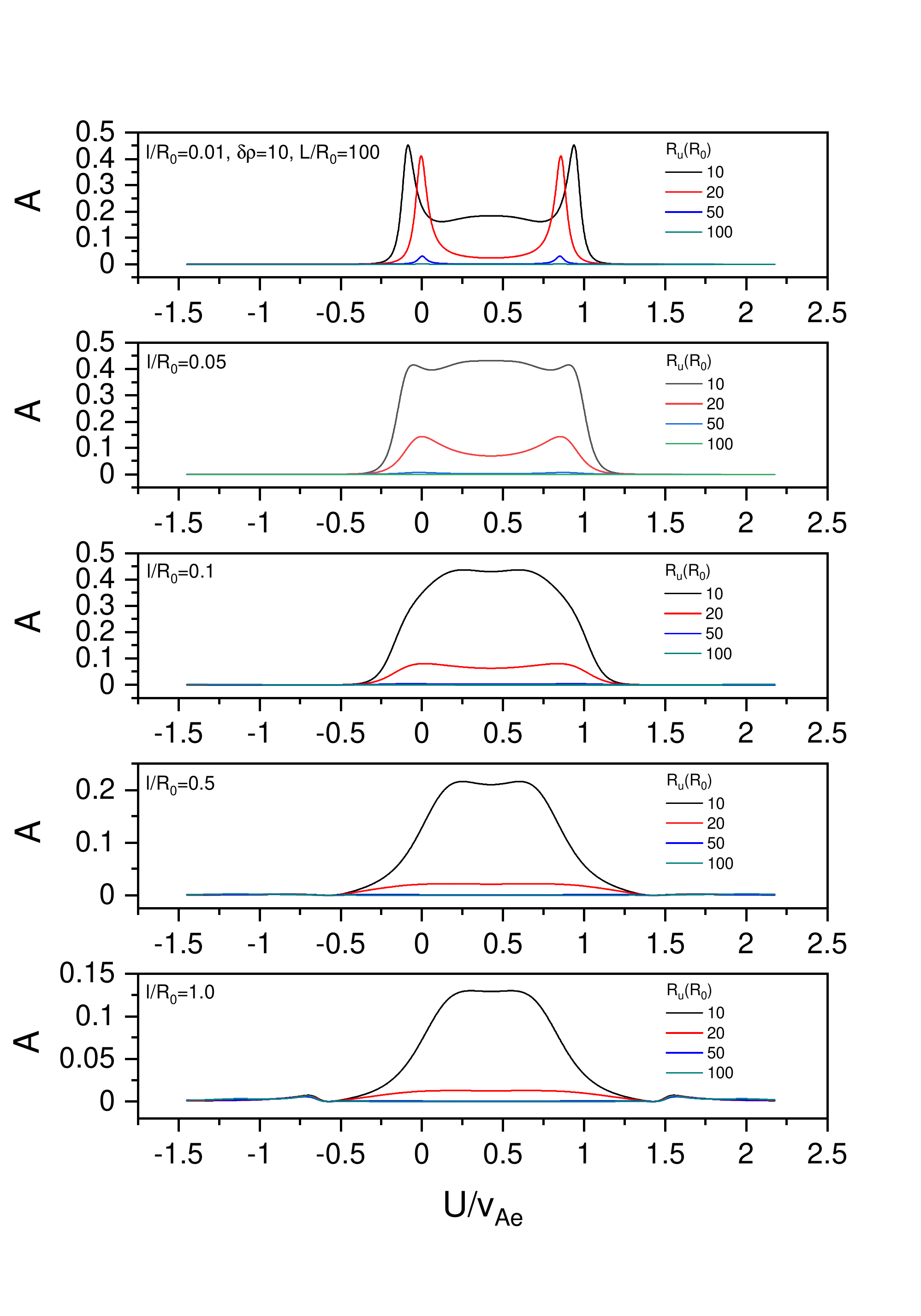}
\caption{\label{f8} Similar to Fig.~\ref{f6} except ${L/R_0}=100$. From top to bottom $l/R_0$=0.01, 0.05, 0.1, 0.5, and 1.0. }
\end{figure}
{These results show that for the long coronal loops resonant absorption can change greatly within a small variation of flow speed when the thickness of the transitional layer is small. On the other hand, for the short coronal loops resonant absorption can be strong for a sufficiently high flow speed in a wide range of flow regime when the transitional layer is thick.} In each case, the absorption is sensitively dependent on the variation of the flow speed.

In previous figures, it is shown that as ${R}_u/R_0$ increases from 20 to 50, $A$ decreases in {regime} I while it increases in {regime} II. Here we investigate in more detail the ${R}_u/R_0$ dependent behavior of $A$ for each ${L/R_0}$ that changes from 10 to 300. Fig.~\ref{f6} describes the results for ${L/R_0}=10$ where ${R}_u/R_0=$10, 20, 50, and 100 for each ${l/R_0}$ that varies from 0.01 to 1. $A$ in {regime} II is dominant and strong in proportion to ${l/R_0}$ when ${l/R_0}>0.5$ regardless of the value of ${R}_u/R_0$. The increase of ${R}_u/R_0$ strengthens the oscillation features. It is noticeable that even if ${l/R_0}$ and ${R}_u/R_0$ are small, $A$ in {regime} I is very small.

For ${L/R_0}=50$, as described in Fig.~\ref{f7}, the above features weaken. When ${R}_u/R_0=10$, $A$ in {regime} II is reduced and no longer effective for all ${l/R_0}$, while $A$ in {regime} I is largely enhanced for small ${l/R_0}$. As ${R}_u/R_0$ increases $A$ in {regime} I recedes and, after ${R}_u/R_0$ is over 20, becomes negligibly small for all ${l/R_0}$.

Fig.~\ref{f8} presents that for ${L/R_0}=100$, $A$ in {regime} II is ineffective while $A$ in {regime} I is dominant. For ${l/R_0}<0.1$, $A$ in {regime} I is sufficiently strong when ${R}_u/R_0=10$. As ${l/R_0}$ increases, two sharp absorption peaks disappear and a new absorption hill (bump-like curve) appears, {which spans a range of $\sim1{U/v_{Ae}}$ around ${U}={v}_k$ (cf. Fig.~\ref{f7}).} When ${R}_u/R_0=20$, $A$ is strong only when ${l/R_0}\approx0.01$. When ${R}_u/R_0>20$, resonant absorption is ineffective for all range of ${l/R_0}$ and ${U/v_{Ae}}$.

\begin{figure}[]
\includegraphics[width=0.5\textwidth]{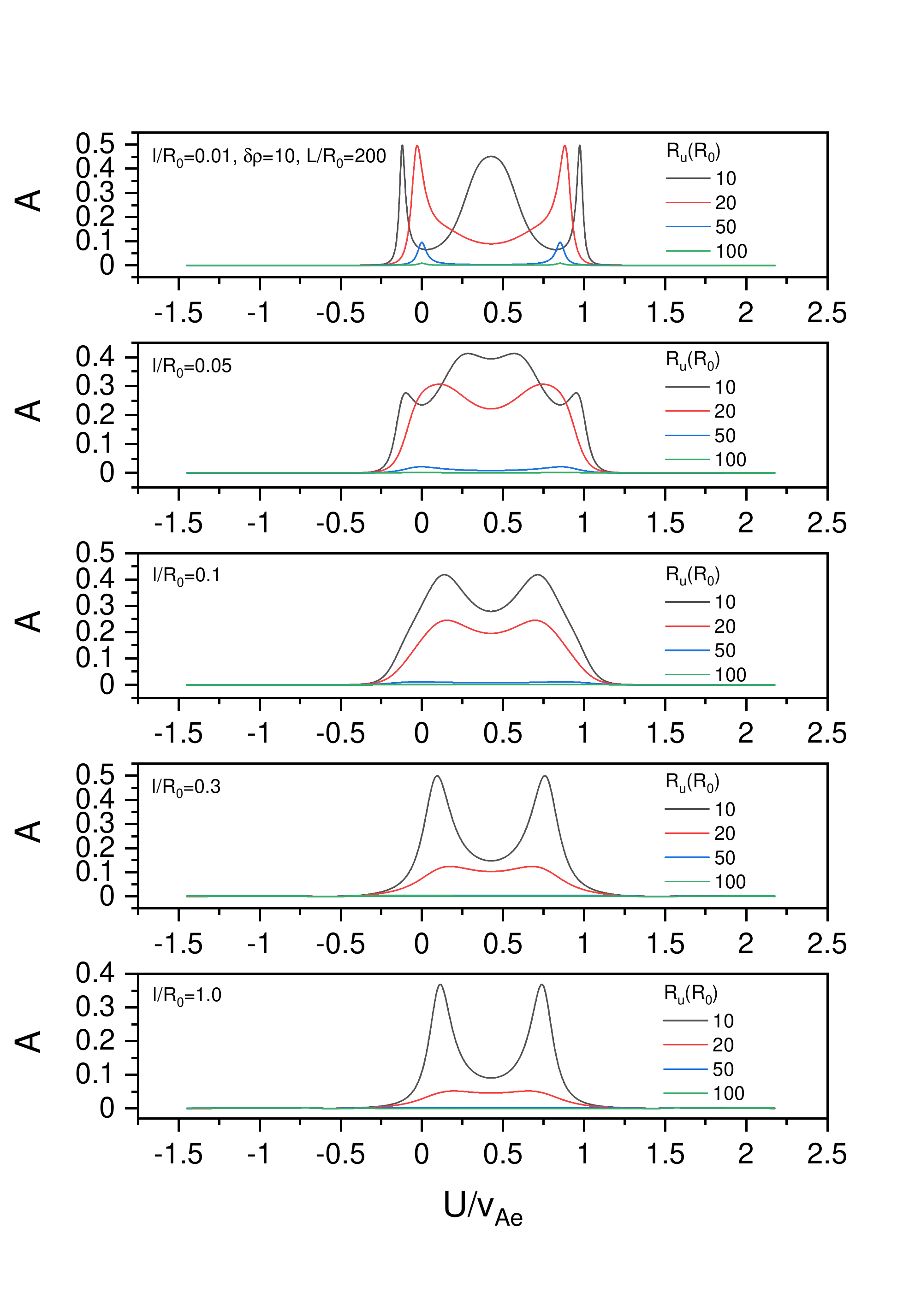}
\caption{\label{f9} Similar to Fig.~\ref{f6} except ${L/R_0}$=200. From top to bottom  ${l/R_0}$=0.01, 0.05, 0.1, 0.3, and 1.0. }
\end{figure}
\begin{figure}[]
\includegraphics[width=0.5\textwidth]{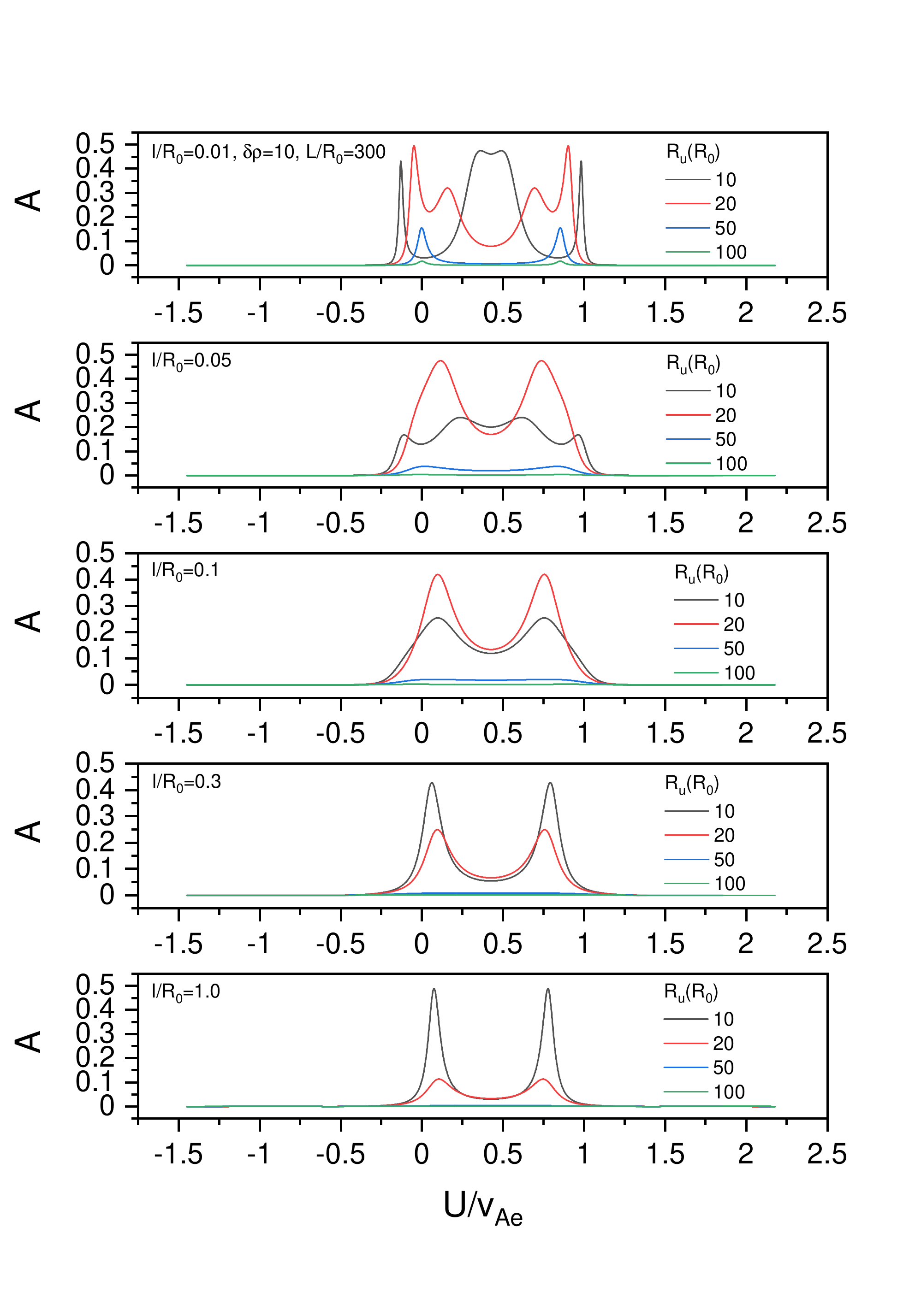}
\caption{\label{f10} Similar to Fig.~\ref{f6} except ${L/R_0}=300$. From top to bottom ${l/R_0}$=0.01, 0.05, 0.1, 0.3, and 1.0. }\end{figure}
In Fig.~\ref{f9}, we plot $A$ vs. ${U/v_{Ae}}$ when ${L/R_0}=200$. It reveals a dramatic change of $A$ for small ${l/R_0}$ when ${R}_u/R_0=10$ (see also Fig.~\ref{f10}).
For ${l/R_0}=0.01$, $A$ has a sharp peak at ${U/v_{Ae}}=-0.12$ (the other responsive peak is at ${U/v_{Ae}}=0.9728$), reaching a maximum value, 0.498, from 0.067 at ${U}=0$. As ${R}_u/R_0$ increases, the absorption peak at ${U/v_{Ae}}=-0.12$ shifts toward ${U}=0$ with decrement and broadening. It is notable that the absorption peak that appears at ${U}={v}_k$ when ${R}_u/R_0=10$ disappears when ${R}_u/R_0$ is over 20.
For ${l/R_0}=0.05$, the absorption curve becomes more complex, having two peaks near ${U}={v}_k$ and a local dip at ${U}={v}_k$. The four absorption peaks converge into two peaks with the increment of ${l/R_0}$.
For ${l/R_0}\geq0.1$, $A$ gradually decreases as ${R}_u/R_0$ increases, maintaining its tendency.
When ${{R}_u/R_0}$ is sufficiently large (here ${R}_u/R_0>20$) $A$ is inversely proportional to ${R}_u/R_0$ regardless of ${l/R_0}$. When ${R}_u/R_0>50$ there is almost no resonant absorption for all ${l/R_0}$. The result for ${{L/R_0}}=50$ can be thought as the intermediate state between ${{L/R_0}}=10$ and ${{L/R_0}}=100$ cases. The range of ${U/v_{Ae}}$ and ${l/R_0}$ for high resonant absorption in {regime} I tends to increase as ${L/R_0}$ increases.

When ${L/R_0}$ is over 100, the hill-like absorption curve that appeared in Figs.~\ref{f7} and \ref{f8} develops into two sharp peaks as shown in Figs.~\ref{f9} and~\ref{f10}. The increment of ${L/R_0}$ also tends to increase the absorption and to extend the range of ${l/R_0}$ and ${R}_u/R_0$ for high absorption. Fig.~\ref{f10} also shows an interesting signature. For some range of ${l/R_0}$, $A$ can be rather enhanced when ${R}_u/R_0$ increases from 10 to 20 (see the second and third rows). For large ${L/R_0}$ the absorption behavior can be very complex when ${l/R_0}$ and ${R}_u/R_0$ are sufficiently small. These results emphasize that the {spatial extent} of the flow region is also a crucial parameter for resonant absorptions when the loop length (or axial wave{length}) is large.

It is now apparent that resonant absorption is significantly and sensitively dependent on the {spatial extent} of the flow region (${R}_u/R_0$), the flow speed (${U/v_{Ae}}$), and the direction of the flow. These effects are conspicuous for long loop lengths (small $k_z$) when ${R}_u/R_0$ is sufficiently small.
For high flow speeds resonant absorption can be effective for the short loop lengths regardless of the value of ${R}_u/R_0$ when ${l/R_0}$ is large.

Finally, we apply our theory to the observation {of kink oscillation} in~\cite{Aschwanden2011}, which reported that $v_{Ae}=1940kms^{-1}$, $B_0=4G$, $\delta\rho=12$, ${L/R_0}=72$, and $R_D/R_0=128$, where $R_D$ is the distance between the flare site and the loop considered. It is obtained that $A$ is below 0.025 for all ${l/R_0}$ when $-1000kms^{-1}<U<1500km^{-1}$. Unfortunately, the shear flow can not explain the observed loop oscillations for $\omega=\omega_k$ because the values of $R_D/R_0$ and ${L/R_0}$ are large. {For large ${R_D/R_0}$, selective excitation and effective resonant absorption are possible only when the flow speed is unusually very high and the loop length is small. Selective excitation due to background flow is effective when $R_D/R_0$ is small.} One possible explanation {for the oscillation} is that the phase speed of the oscillation is higher than the cutoff~\citep{Yu2016b}, which is supported by the fact that the oscillation, which has no damping for a long time, contradicts the rapid damping predicted by resonant absorption. Another possibility is that a strong fast shock generated at the flare site induces the loop oscillations. The high amplitude of the wave may enhance the wave propagation to the loop and excite the loop oscillations, which needs nonlinear studies.

\section{Conclusions}
\label{sec4}

We studied the effects of intermediate shear flow on resonant absorption of coronal loop kink oscillations where an external wave is incident on the loop. Resonant absorption was known to significantly depend on the loop length, density profile (thickness of the transitional layer and density contrast), and wave frequency~\citep[e.g.][]{Ruderman2002,VanDoorsselaere2004,Arregui2007,Soler2013,Yu2016b,Yu2017a,Yu2017b,Yu2019}. The shear flow inside the loop is also an important factor to it~\citep{Goossens1992,Ruderman2010,Ruderman2019,Soler2019,Sadeghi2021}. The presence of shear flow region outside the loop is now added as another crucial factor for resonant absorption if the loop oscillation is excited externally by a wave source like flares. We showed that for long coronal loops resonant absorption can be severely enhanced or reduced with a small change of flow speed in the direction parallel or antiparallel to the axial wave vector. When the loop length is short and the transitional layer is thick, high flow speed is needed to have considerable resonant absorption. Resonant absorption is also sensitive to the {spatial extent} of the flow region.

Since the strong (weak) resonant absorption means the high (low) transmission or transport of the wave energy into the loop, the results mean that the shear flow can control the excitation of the loop kink oscillations. This further implies that even if the two loops, which are adjacent to each other, have the same structure and the same wave modes are excited, one loop does show oscillation while the other loop does not depending on the profile of the surrounding shear flow.
It is desirable to see whether these effects are valid in 2D/3D simulations~\citep[e.g.][]{Luna2008,Luna2009,Karampelas2020,Magyar2020}.

Our results also emphasize the role of the external Alfv\'{e}n speed ($v_{Ae}$) and phase speed of the wave (here $v_k$). These parameters determine the range and the boundary of two absorption {regimes}, which are distinguished by whether the potential $\delta{V}$ is below or above 1, that yields different absorption behaviors for each {regime}. The range of flow speed for the substantial change of resonant absorption crucially depends on $v_{Ae}$. The symmetry of resonant absorption with respect to $v_{k}$ is also worthy of notice.

We considered a rectangular profile for the shear flow which can, however, have more complexities like (quasi-)periodic or random structures~\citep[e.g.][]{Murawski2000}. These systems may increase the backscattering of the incident wave, reducing the wave transport to the loop.  The continuous spatial variation of the flow may induce Alfv\'{e}n resonance in the shear flow region if it satisfies the condition $\Omega=\pm\omega_A$~\citep[][]{Goossens1992,Csik1998}. This effect may further contribute to the dissipation of the wave energy during the propagation.
From the potential, Eq.~(\ref{eq:7}), it is expected that the shear flow plays a similar role to the density and the absorption behavior becomes more complicated ~\citep[e.g.][]{Murawski2001,Yu2010,Yu2013,Yu2016a,Yuan2015}.

\acknowledgments
The author thanks the referee for constructive comments.
 This research was supported by Basic Science Research Program through the National Research Foundation of Korea(NRF) funded by the Ministry of Education (No. 2020R1I1A1A01066610).

\bibliographystyle{aasjournal}

\end{document}